\documentclass[%
 reprint,
superscriptaddress,
 amsmath,amssymb,
 aps,
floatfix,
]{revtex4-2}

\usepackage[table]{xcolor}
\usepackage{xcolor} 
\usepackage{graphicx}
\usepackage{dcolumn}
\usepackage{bm}
\usepackage{physics}
\usepackage{appendix}
\usepackage{float}
\usepackage{hyperref}
\usepackage[percent]{overpic}

\begin{document}

\preprint{APS/123-QED}

\title{Auger Spectroscopy via Generative Quantum Eigensolver: A Quantum Approach to Molecular Excitations}
\date{\today}

\begin{abstract}
Auger electron spectroscopy, a way of characterizing electronic structure through core-level decay processes, is widely used in materials characterization; however direct calculation from molecular geometry requires accurate treatment of many excited states, posing a challenge for classical methods.
We present a hybrid quantum-classical workflow for calculating Auger spectra that combines the generative quantum eigensolver (GQE) for ground-state preparation, the quantum self-consistent equation-of-motion method for excited-state calculations, and the one-centre approximation for Auger transition rates.
GQE uses a GPT-2 model to generate quantum circuits for ground-state optimization, allowing our workflow to benefit from HPC parallelization and GPU-acceleration for favourable scaling with system size.
We demonstrate the validity of our workflow by calculating the Auger spectrum of water with the STO-3G basis set and demonstrating qualitative and quantitative agreement with spectra obtained using completely classical full configuration interaction calculations, from the computational literature, and from the experimental literature.
We also find that for water, substituting the variational quantum eigensolver (VQE) for GQE results in near-identical spectra, but that the ground state estimator generated by GQE contains about half the total gate count as that generated by VQE.

\end{abstract}

\author{Kimberlee Keithley}
\thanks{These authors contributed equally to this work.}
\affiliation{Mitsubishi Chemical Corporation, Science \& Innovation Center,
1000, Kamoshida-cho, Aoba-ku, Yokohama 227-8502, Japan}
\affiliation{Quantum Computing Center, Keio University, Hiyoshi 3-14-1,
Kohoku, Yokohama 223-8522, Japan}

\author{Shunsuke Yamamoto}
\thanks{These authors contributed equally to this work.}
\affiliation{Mizuho Research \& Technologies, Ltd., 2-3 Kanda-Nishikicho, Chiyoda-ku, Tokyo 101-8443, Japan}

\author{Ryota Kenmoku}
\affiliation{Quantum Computing Center, Keio University, Hiyoshi 3-14-1,
Kohoku, Yokohama 223-8522, Japan}

\author{Ikko Hamamura}
\affiliation{
NVIDIA, ATT EAST 12F, 2-11-7 Akasaka, Minato-ku, Tokyo 107-0052, Japan
}

\author{Kouhei Nakaji}
\affiliation{
NVIDIA, Santa Clara, California, USA
}

\author{Shu Kanno}
\affiliation{Mitsubishi Chemical Corporation, Science \& Innovation Center,
1000, Kamoshida-cho, Aoba-ku, Yokohama 227-8502, Japan}
\affiliation{Quantum Computing Center, Keio University, Hiyoshi 3-14-1,
Kohoku, Yokohama 223-8522, Japan}

\author{Takao Kobayashi}
\affiliation{Mitsubishi Chemical Corporation, Science \& Innovation Center,
1000, Kamoshida-cho, Aoba-ku, Yokohama 227-8502, Japan}

\author{Qi Gao}
\affiliation{Mitsubishi Chemical Corporation, Science \& Innovation Center,
1000, Kamoshida-cho, Aoba-ku, Yokohama 227-8502, Japan}
\affiliation{Quantum Computing Center, Keio University, Hiyoshi 3-14-1,
Kohoku, Yokohama 223-8522, Japan}

\author{Shumpei Uno}%
\affiliation{Mizuho Research \& Technologies, Ltd., 2-3 Kanda-Nishikicho, Chiyoda-ku, Tokyo 101-8443, Japan}

\author{Kohei Oshio}
\affiliation{Mizuho Research \& Technologies, Ltd., 2-3 Kanda-Nishikicho, Chiyoda-ku, Tokyo 101-8443, Japan}
\affiliation{Quantum Computing Center, Keio University, Hiyoshi 3-14-1,
Kohoku, Yokohama 223-8522, Japan}

\author{Naoki Watanabe}
\affiliation{Mizuho Research \& Technologies, Ltd., 2-3 Kanda-Nishikicho, Chiyoda-ku, Tokyo 101-8443, Japan}

\author{Takeshi Sato}
\affiliation{Department of Nuclear Engineering and Management, Graduate School of Engineering, The University of Tokyo, 7-3-1 Hongo, Bunkyo-ku, Tokyo 113-8656, Japan}

\author{Naoki Yamamoto}
\affiliation{Quantum Computing Center, Keio University, Hiyoshi 3-14-1,
Kohoku, Yokohama 223-8522, Japan}

\author{Shunya Minami}
\affiliation{Global R\&D Center for Business by Quantum-AI Technology, National Institute of Advanced Industrial Science and Technology (AIST), 1-1-1 Umezono, Tsukuba, Ibaraki, Japan}

\author{Yohichi Suzuki}
\affiliation{Global R\&D Center for Business by Quantum-AI Technology, National Institute of Advanced Industrial Science and Technology (AIST), 1-1-1 Umezono, Tsukuba, Ibaraki, Japan}


\author{Yuma Nakamura}
\affiliation{Department of Computer Science, University of Toronto, 40 St George St., Toronto, ON M5S 2E4, Canada}
\affiliation{Vector Institute for Artificial Intelligence, W1140-108 College St., Schwartz Reisman Innovation Campus, Toronto, ON M5G 0C6, Canada}
\affiliation{Acceleration Consortium, 700 University Ave., Toronto, ON M7A 2S4, Canada}
\author{Jorge A. Campos-Gonzalez-Angulo}
\affiliation{Department of Chemistry, University of Toronto,  80 St. George St., Toronto, ON M5S 3H6, Canada}

\author{Mohammad Ghazi Vakili}
\affiliation{Department of Computer Science, University of Toronto, 40 St George St., Toronto, ON M5S 2E4, Canada}
\affiliation{Department of Chemistry, University of Toronto,  80 St. George St., Toronto, ON M5S 3H6, Canada}

\author{Al\'an Aspuru-Guzik}
\affiliation{Department of Chemistry, University of Toronto,  80 St. George St., Toronto, ON M5S 3H6, Canada}
\affiliation{Department of Computer Science, University of Toronto, 40 St George St., Toronto, ON M5S 2E4, Canada}
\affiliation{Vector Institute for Artificial Intelligence, W1140-108 College St., Schwartz Reisman Innovation Campus, Toronto, ON M5G 0C6, Canada}
\affiliation{Acceleration Consortium, 700 University Ave., Toronto, ON M7A 2S4, Canada}
\affiliation{Department of Chemical Engineering \& Applied Chemistry, University of Toronto, 200 College St., Toronto, ON M5S 3E5, Canada}
\affiliation{Department of Materials Science \& Engineering, University of Toronto, 184 College St., Toronto, ON M5S 3E4, Canada}
\affiliation{Institute of Medical Science,  University of Toronto, 1 King's College Circle, Medical Sciences Building, Room 2374, Toronto, ON M5S 1A8, Canada}
\affiliation{NVIDIA, 431 King St. W \#6th, Toronto, ON M5V 1K4, Canada}
\affiliation{Canadian Institute for Advanced Research (CIFAR), 661 University Ave., Toronto,
ON M5G 1M1, Canada}
\maketitle

\section{Introduction}
Computational prediction of material properties can accelerate the timeline and reduce the cost of developing new materials~\cite{curtarolo_high-throughput_2013, hautier_finding_2019}.
In electronics manufacturing, computational design has proven useful in engineering photosensitive materials for use in photolithographic processes~\cite{manikandan_computational_2024, lian_molecular_2025}.
To meet increasing demands for smaller devices, extreme ultraviolet (EUV) lithography has become increasingly popular~\cite{schofield_roadmap_2025}.
In environments with EUV radiation, the decay of core-electron excitations heavily influences photoresist performance, making understanding these decay pathways essential for engineering next-generation materials~\cite{closser_importance_2017}.
These excitation and decay processes are measured experimentally by Auger spectroscopy.
Auger spectra are also commonly used in other industrial processes such as in development of solar devices~\cite{ricciari_auger_2015} or thin film deposition techniques~\cite{li_auger_2022}.

To obtain Auger spectra computationally, one must accurately determine both the ground and excited electronic states of the system.
Since Auger spectra involve core-ionized and doubly-ionized states, the number of excited states that must be resolved is large.
Exact diagonalization via full configuration interaction (FCI) scales exponentially with the number of orbitals and becomes intractable beyond approximately 20 spatial orbitals~\cite{sun_recent_2020, gao_distributed_2024}. Moreover, approximate methods such as complete active space self-consistent field and restricted active space self-consistent field (RASSCF) require careful selection of active spaces~\cite{tenorio_multireference_2022, rankine_progress_2021}.
Although coupled-cluster-based approaches such as equation-of-motion coupled-cluster singles and doubles method combined with the Feshbach--Fano formalism~\cite{Skomorowski2021AugerI, Skomorowski2021AugerII} or the one-centre approximation (OCA)~\cite{Skomorowski2025OCA} have demonstrated accurate Auger decay rates for small molecules, these methods remain computationally demanding.

Quantum computers are may offer significant advantages over classical computers in performing quantum chemical calculations due to their ability to natively represent quantum information~\cite{feynman_simulating_nodate}.
Variational quantum eigensolvers (VQEs) are a versatile and popular class of quantum-hybrid algorithms with applications in chemistry~\cite{peruzzo_variational_2014} as well as a wide variety of fields from combinatorial mathematics to condensed matter~\cite{cerezo_variational_2021, fedorov_vqe_2022}.
In VQE, a quantum circuit that evaluates the expectation value of a Hamiltonian with respect to an eigenstate (in this case, the ground state) is prepared iteratively, parametrized by a set of adjustable continuous input gate parameters.
However, VQE suffers from a few key drawbacks that prevent it from scaling to large systems.
The first is that algorithm performance depends acutely on the choice of ansatz~\cite{mao_towards_2024, kandala_hardware-efficient_2017}.
The second is that variational quantum algorithms suffer from a scaling problem known as the barren plateau~\cite{larocca_barren_2025, mcclean_barren_2018}: as the number of qubits increases, the gradient of the cost function decreases exponentially, and the classical optimizer is unable to progress.

The generative quantum eigensolver (GQE) algorithm~\cite{nakaji_generative_2024} leverages a generative pre-trained transformer (GPT) model~\cite{Radford2019LanguageMA} inspired by the success of GPT models applied to large language models (LLMs)~\cite{vaswani2017attention} to generate a probability distribution of circuits with desirable properties.
The use of classical GPT models allows GQE to benefit from HPC parallelization and GPU tensor arithmetic acceleration.
GQE might also be compatible with frameworks for molecule-to-molecule transfer training, which will enable the treatment of larger molecules with models trained on smaller ones~\cite{thiede2026coupled}.
Additionally, the ability of GQE to generate entire circuits rather than just input gate parameters, as with VQE~\cite{verdon_learning_2019, kim_preparation_2023, yang_maximizing_2024}, may avoid the barren plateau problem\cite{nakaji_generative_2024} or offer reduced circuit depth.

In this work, we develop a computational workflow to calculate Auger spectra from ground-state GQE calculations.
Although NISQ algorithms for calculating excited-state spectra have been proposed and tested on small molecules~\cite{shirai_calculation_2022, cadi_tazi_folded_2024}, this is, to the authors' knowledge, the first use of GQE to calculate excited state spectra.
We use the quantum self-consistent equation-of-motion (q-sc-EOM) method developed by Asthana et al.~\cite{asthana_quantum_2023} to obtain excited states from GQE ground-state calculations, and the OCA to calculate Auger spectra from those states.
We apply this workflow to calculate the Auger spectrum of a water molecule using the STO-3G basis set.

\section{Methods}\label{sec:methods}
\begin{figure*}[htbp]
    \includegraphics[width=\textwidth]{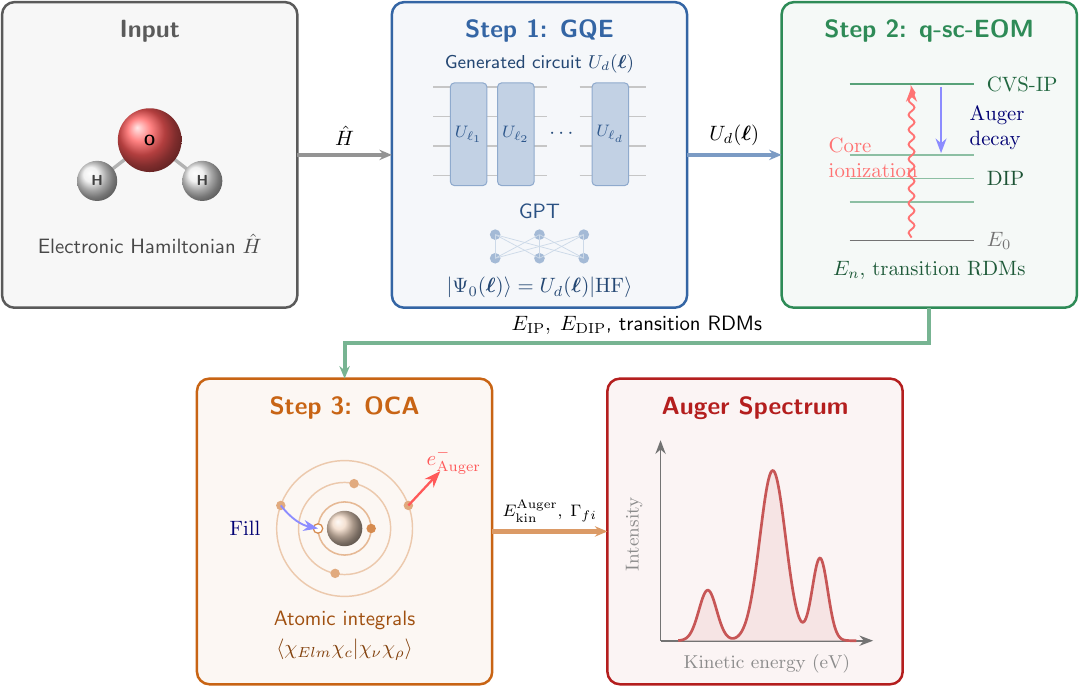}
    \caption{Graphical summary of the Auger workflow used in this study. First, the ground-state reference is prepared with the GQE~\cite{nakaji_generative_2024}. Next, q-sc-EOM provides core-ionized and doubly ionized energies ($E_{\mathrm{IP}}$, $E_{\mathrm{DIP}}$) and transition reduced density matrices~\cite{asthana_quantum_2023}. Finally, the OCA~\cite{tenorio_multireference_2022} combines these quantities with atomic integrals to produce the Auger spectrum.}
    \label{fig:workflow}
\end{figure*}

To obtain Auger spectra, we first calculate the ground-state energy of each molecule using the GQE algorithm~\cite{nakaji_generative_2024}, followed by application of the q-sc-EOM method~\cite{asthana_quantum_2023} to generate the excitation spectra.
All quantum-circuit evaluations in both the GQE and q-sc-EOM steps are performed using the CUDA-Q~\cite{cudaq_software} state-vector simulator without noise.
From these results, we use the OCA~\cite{siegbahn_auger_1975, agren_theory_1992, fink_autoionization_1995, fink_autoionization_1997, tenorio_multireference_2022} to calculate Auger spectra.
Our workflow uses a frozen-core strategy in the ground-state reference and applies a core-valence separation (CVS) constraint in the ionization manifolds.
This design is conceptually aligned with frozen-core CVS-EOM-CC approaches developed for core-excited and core-ionized states~\cite{vidal_new_2019}.
In our quantum setting, the frozen-core approximation is implemented by excluding core spin orbitals from the qubit register during ground-state preparation and reintroducing them at the q-sc-EOM stage.
Each spin orbital is mapped to a single qubit via the Jordan--Wigner transformation~\cite{Jordan1928-hf}, and all operators built from fermionic creation and annihilation operators are expressed as qubit (Pauli) operators accordingly.
We briefly detail the theory and algorithms below.
The Auger branch of this workflow is summarized in Fig.~\ref{fig:workflow}.

\subsection{Generative Quantum Eigensolver}\label{subsec:gqe}

GQE works analogously to large language models, but where the output of the model is quantum circuits rather than language documents.
To achieve this, GQE combines GPT2LMHeadModel from the Hugging Face Transformers package, the publicly available version of the model underlying early versions of the popular chat bot ChatGPT, originally developed by Radford et al.~\cite{Radford2019LanguageMA}, with a chemically-inspired quantum operator pool.
The end result is a quantum circuit, rather than a document.
As with GPT LLMs, GQE generates a probability distribution over a set of tokens; however rather than tokens being taken from parts of given language (words, punctuation, etc.), the tokens are sequences of quantum gates.
Also like GPT LLMs, the generative model learns to order tokens to achieve a desirable outcome.
In the case of language models, it is desirable to order tokens to follow grammatical rules; in the case of GQE we desire to estimate the ground state energy, or equivalently, to minimize the Boltzmann-weighted expectation of the Hamiltonian associated with a chemical system.

The depth parameter $d$ controls the number of tokens (words) per circuit (document).
The operator pool $\{U_\ell\}_{\ell=1}^{L}$ is a set of time-evolved Pauli operators from the unitary coupled-cluster singles and doubles (UCCSD) ansatz, $\{e^{i \hat{P}_\ell t_\ell}\}_{\ell=1}^{L}$.
Here, $L$ denotes the number of operators in the pool.
$\hat{P}_\ell$ are the Pauli strings of length $n\_qubits$ derived from the UCCSD ansatz, where $n\_qubits$ is the number of qubits used in the molecular model.
The $t_\ell$ are generated from the discrete time pool $\{\pm2^k / 160\}_{k=-1}^4$.
Including additional terms in the time pool did not yield significant changes in accuracy for a given document depth.

A GQE circuit of depth $d$ is constructed as an ordered product of $d$ operators drawn from the pool $\{U_\ell\}$:
\begin{equation}\label{eqn:gqe_circuit}
\begin{split}
U_d(\boldsymbol{\ell}) &= U_{\ell_d}\cdots U_{\ell_2}U_{\ell_1},\\
\ell_m &\in \{1,\ldots,L\},\quad m=1,\ldots,d.
\end{split}
\end{equation}
The reference state is obtained by applying $U_d(\boldsymbol{\ell})$ to the Hartree--Fock determinant:
\begin{equation}\label{eqn:gqe_state}
| \Psi_0(\boldsymbol{\ell}) \rangle = U_d(\boldsymbol{\ell}) | \mathrm{HF} \rangle.
\end{equation}

The model is trained to minimize the Boltzmann weighted expectation value of the Hamiltonian of interest via reinforcement learning using group relative policy optimization~\cite{shao2024deepseekmathpushinglimitsmathematical} as the loss function.
By the variational principle, this procedure gives us an estimate of the ground state energy.
A summary of the analogy between GPT models trained on language and GQE is shown in Tab.~\ref{tab:gqe-llm}

\begin{table}
    \caption{A summary of the analogy between GQE and GPT models trained on language.}
    \label{tab:gqe-llm}
    \begin{tabular}{ccc}
        \hline
        & LLM & GQE \\
        \hline
        Product & Document & Circuit \\
        Tokens & Words & $e^{i \hat{P}_\ell t_\ell}$ \\
        Learning objective & Grammar & $\mathrm{min\,} e^{\beta E}$ \\
        \hline
    \end{tabular}
\end{table}

Details of the training algorithm are discussed below.
The repetition penalty for the GPT-2 model is set to 1.2.
Each training step generates a batch of $batch\_size$ circuits.
The fitness of the entire batch is evaluated by ranking each sampled circuit relative to other circuits in the same batch.
We label the circuit with the lowest energy (Boltzmann-weighted expectation value of $\hat{H}$) the sole winner within a training batch, and all other circuits are classified as losers for the purpose of backpropagation.
PyTorch's AdamW~\cite{loshchilov2019decoupledweightdecayregularization} optimizer is used to update the learning parameters of GPT-2.
Every $epoch\_size$ training steps constitutes an epoch, at the end of which the training buffer which holds previously generated circuits and their energies, is updated.
This buffer of circuit and energy pairs is used in gradient calculation, cutting down on the number of circuit evaluations in each training cycle.

The number of samples per batch ($batch\_size$) and buffer size are set to 50 by default.
The number of steps per training epoch ($epoch\_size$) is set to 30.
Unless otherwise stated, all simulations use the default parameters listed above.
A detailed explanation of GQE is available in Ref.~\cite{nakaji_generative_2024}.

\subsection{Quantum Self-Consistent Equation of Motion}\label{subsec:qsceom}
To obtain excited-state and ionized-state properties from GQE results, we employ the q-sc-EOM method developed by Asthana et al.~\cite{asthana_quantum_2023}, building on the quantum equation-of-motion (qEOM) approach of Ollitrault et al.~\cite{ollitrault_quantum_2020}.
The q-sc-EOM method is part of a class of approaches known as quantum subspace methods~\cite{motta_subspace_2024, barison_sqd_excited_2025, yu_sqd_krylov_2025, nakamura_adaptive_2024}, which project the electronic Hamiltonian onto a subspace spanned by states prepared on the device.

The q-sc-EOM procedure constructs a subspace spanned by basis states formed by applying the unitary operator $U$ obtained by GQE to states generated from the Hartree--Fock determinant, $\hat{G}_u | \mathrm{HF} \rangle$:
\begin{equation}\label{eqn:basis}
| \psi_u \rangle = U \, \hat{G}_u | \mathrm{HF} \rangle,
\end{equation}
where $\hat{G}_u$ is a product of fermionic creation and annihilation operators.
Here, indices $i, j, k, \ldots$ denote occupied spin orbitals and $a, b, \ldots$ virtual spin orbitals.
Different spectroscopic channels are accessed by choosing appropriate operator types: particle-number-conserving excitations ($\hat{G}_u \in \{ \hat{a}^\dagger_p \hat{a}_q, \, \hat{a}^\dagger_p \hat{a}^\dagger_q \hat{a}_r \hat{a}_s \}$) for electron excitation, single-ionization operators ($\hat{G}_u \in \{ \hat{a}_p, \, \hat{a}^\dagger_p \hat{a}_q \hat{a}_r \}$) for ionization potential (IP), and double-ionization operators ($\hat{G}_u \in \{ \hat{a}_p \hat{a}_q, \, \hat{a}^\dagger_r \hat{a}_p \hat{a}_q \hat{a}_s \}$) for double ionization potential (DIP).

The key quantity is the $\mathbf{M}$ matrix with elements
\begin{equation}\label{eqn:mij}
M_{uv} = \langle \psi_u | \hat{H} | \psi_v \rangle = \langle \mathrm{HF} | \hat{G}_u^\dagger \, U^\dagger \, \hat{H} \, U \, \hat{G}_v | \mathrm{HF} \rangle.
\end{equation}
When point-group symmetry is available, the excitation operators $\{\hat{G}_u\}$ are further classified by irreducible representation (irrep) based on the direct product of the orbital irreps of the involved spin orbitals.
This classification affords the block diagonalization of the $\mathbf{M}$ matrix, $\mathbf{M} = \bigoplus_{\Gamma} \mathbf{M}^{(\Gamma)}$, reducing the number of off-diagonal measurements from $O(N_{\mathrm{exc}}^2)$ to $\sum_\Gamma O(N_\Gamma^2)$, where $N_{\mathrm{exc}}$ is the total number of excitation operators and $N_\Gamma$ is the number of excitations in irrep $\Gamma$.
The diagonalization, transition reduced density matrix (RDM) computation, and spectral analysis then proceed independently for each irrep block.

The excited-state energies $E_n$ and expansion coefficients $\mathbf{c}^{(n)}$ are obtained by solving the eigenvalue problem
\begin{equation}\label{eqn:eig}
\mathbf{M} \, \mathbf{c}^{(n)} = E_n \, \mathbf{c}^{(n)},
\end{equation}
defining the approximate eigenstates $| \Psi_n \rangle = \sum_u c_u^{(n)} | \psi_u \rangle$.

Diagonal elements $M_{uu}$ are measured on the quantum device by preparing the state $U \, \hat{G}_u | \mathrm{HF} \rangle$ and evaluating the Hamiltonian expectation value.
Off-diagonal elements are obtained via a superposition technique: by constructing the superposition state
\begin{equation}\label{eqn:superposition}
| \Phi_{uv}^{(\phi)} \rangle = \frac{1}{\sqrt{2}} \left( \hat{G}_u + e^{i\phi} \hat{G}_v \right) | \mathrm{HF} \rangle
\end{equation}
and applying the unitary operator $U$ before measuring the Hamiltonian expectation value
\begin{equation}\label{eqn:e_phi}
\begin{split}
E_{uv}^{(\phi)} = \frac{1}{2} \Big\langle \mathrm{HF} \Big| \left( \hat{G}_u + e^{i\phi} \hat{G}_v \right)^\dagger U^\dagger \, \hat{H} \, U \\
\times \left( \hat{G}_u + e^{i\phi} \hat{G}_v \right) \Big| \mathrm{HF} \Big\rangle
\end{split}
\end{equation}
at phase angles $\phi = 0$ and $\phi = \pi/2$, one extracts the real and imaginary parts of the off-diagonal elements as
\begin{align}
\mathrm{Re}(M_{uv}) &= E_{uv}^{(0)} - \tfrac{1}{2}\left( M_{uu} + M_{vv} \right), \label{eqn:re_mij} \\
\mathrm{Im}(M_{uv}) &= -E_{uv}^{(\pi/2)} + \tfrac{1}{2}\left( M_{uu} + M_{vv} \right). \label{eqn:im_mij}
\end{align}
This approach avoids the generalized eigenvalue problem of the original qEOM formulation~\cite{ollitrault_quantum_2020}. Additionally, the use of self-consistent operators in q-sc-EOM automatically satisfies the vacuum annihilation condition~\cite{asthana_quantum_2023}, improving the accuracy of the calculated spectra.

Because the approximate reference and truncated q-sc-EOM subspace do not guarantee perfectly spin-pure DIP states, especially for near-degenerate levels, residual spin mixing can occur.
To mitigate this effect, we apply an additional classical post-processing step in the DIP channel.
For each irrep block, we evaluate the DIP $S^2$ and $\mathbf{M}$ matrices, partition the $S^2$ eigenvectors into approximate spin sectors using the nearest ideal odd-multiplicity $S(S+1)$ value, and diagonalize the projected $\mathbf{M}$ matrix within each sector.

To compute spectral properties from the q-sc-EOM eigenstates, we require transition reduced density matrices between different states~\cite{kumar_quantum_2023}. Since the relevant operators are in general non-Hermitian, we decompose a generic operator $\hat{T}$ into two Hermitian parts,
\begin{equation}\label{eqn:herm_decomp}
\hat{T}_{\mathrm{re}} = \tfrac{1}{2}\left( \hat{T} + \hat{T}^\dagger \right), \qquad \hat{T}_{\mathrm{im}} = -\tfrac{i}{2} \left( \hat{T} - \hat{T}^\dagger \right),
\end{equation}
each of which can be measured on the quantum device. The expectation value of the original operator is then recovered as
$
\langle \hat{T} \rangle = \langle \hat{T}_{\mathrm{re}} \rangle + i \, \langle \hat{T}_{\mathrm{im}} \rangle.
$
The matrix element $v_u = \langle \Psi_0 | \hat{T} | \psi_u \rangle$ between the reference state and each basis state $| \psi_u \rangle$ is computed using the same superposition technique described above, pairing the identity operator ($\hat{G}_0 = \hat{\mathbf{1}}$) with each $\hat{G}_u$.
The transition matrix element for the $n$-th eigenstate $| \Psi_n \rangle = \sum_u c_u^{(n)} | \psi_u \rangle$ is then obtained by taking the inner product with the eigenvector coefficients:
\begin{equation}\label{eqn:ref_eigenstate}
\langle \Psi_0 | \hat{T} | \Psi_n \rangle = \sum_u v_u \, c_u^{(n)}.
\end{equation}
More generally, the operator matrix element between two eigenstates is given by
\begin{equation}\label{eqn:operator_eigenbasis}
\langle \Psi_f | \hat{T} | \Psi_n \rangle = \sum_{u, v} \left( c_u^{(f)} \right)^* O_{uv} \, c_v^{(n)},
\end{equation}
where $O_{uv} = \langle \psi_u | \hat{T} | \psi_v \rangle$ is computed using the superposition technique.

For Auger decay, the relevant quantity is the transition RDM between the core-ionized state $| \Psi_I^{N_I} \rangle$ and doubly ionized final states $| \Psi_K^{N_I-1}\rangle$, involving one-electron creation and two-electron annihilation,
\begin{equation}\label{eqn:auger_rdm}
R_{KI;csr} = \langle \Psi_K^{N_I-1} | \hat{a}_c^\dagger \hat{a}_s \hat{a}_r | \Psi_I^{N_I} \rangle,
\end{equation}
which characterizes filling of the core hole in orbital $c$ accompanied by two-electron emission ($r, s$).
This quantity is computed via Eq.~\eqref{eqn:operator_eigenbasis} with $\hat{T} = \hat{a}_c^\dagger \hat{a}_s \hat{a}_r$, where both the initial and final states are excited states obtained from separate q-sc-EOM calculations (IP and DIP channels, respectively).
X-ray absorption spectra can also be calculated using the one-body transition RDM $\gamma_{pq}^{(n)}$ (see Appendix~\ref{app:lih-xas} for details).

\subsection{Auger Decay and One-Centre Approximation}\label{subsec:auger}
In Auger decay, a core-ionized state relaxes by filling the core hole from a valence orbital while ejecting a second valence electron.
The process connects a core-ionized state ($N-1$ electrons) to a doubly ionized final state ($N-2$ electrons).
The Auger electron kinetic energy for a given initial state $I$ and final state $K$ is
\begin{equation}\label{eqn:auger_ekin}
E_{\mathrm{kin}}^{\mathrm{Auger}} = E_{\mathrm{IP}}^{(I)} - E_{\mathrm{DIP}}^{(K)},
\end{equation}
where $E_{\mathrm{IP}}^{(I)}$ is the energy of the core-ionized state obtained from the IP channel of q-sc-EOM, and $E_{\mathrm{DIP}}^{(K)}$ is the energy of the doubly ionized final state obtained from the DIP channel.

The total Auger decay rate for a given initial-final state pair is
\begin{equation}\label{eqn:auger_width}
\Gamma_{KI} = \sum_{l,m} \Gamma_{KI;Elm},
\end{equation}
where $\Gamma_{KI;Elm}$ is the partial decay rate into the continuum channel with angular momentum quantum numbers $(l, m)$.
Each partial rate is calculated via Wentzel's ansatz~\cite{wentzel_1927,mehlhorn_70_1998},
\begin{equation}\label{eqn:auger_decay}
\Gamma_{KI;Elm} \simeq 2\pi \left| B_{KI;Elm} \right|^2,
\end{equation}
where the two-electron matrix element is
\begin{equation}\label{eqn:b_defn}
\begin{split}
B_{KI;Elm} &\equiv \langle \hat{a}_{Elm} \Psi_K^{N_I-1} | \hat{g} | \Psi_I^{N_I} \rangle \\
&= \sum_{crs} \langle \phi_{Elm} \phi_c | \phi_r \phi_s \rangle R_{KI;csr}.
\end{split}
\end{equation}
Here, $\hat{g}$ is the two-electron Coulomb operator, $\phi_{Elm}$ is the continuum orbital of the emitted Auger electron with kinetic energy $E$ and angular momentum quantum numbers $(l, m)$, $\phi_c$ is the core orbital, $\phi_r$ and $\phi_s$ are valence molecular spin orbitals, and $R_{KI;csr}$ is the Auger transition RDM defined in Eq.~\eqref{eqn:auger_rdm}.

The OCA~\cite{siegbahn_auger_1975, agren_theory_1992, fink_autoionization_1995, fink_autoionization_1997, tenorio_multireference_2022} replaces the computationally demanding two-electron integrals in Eq.~\eqref{eqn:b_defn} with atomic integrals localized at the core-hole site $A$:
\begin{equation}\label{eqn:oca_approx}
\begin{split}
\langle \phi_{Elm} \phi_c | \phi_r \phi_s \rangle
&\simeq \sum_{\mu\nu\rho}
  \langle \chi^{A}_{Elm} \chi^{A}_{\mu} | \chi^{A}_{\nu} \chi^{A}_{\rho} \rangle \\
&\quad \times D_{\mu c}\, D_{\nu r}\, D_{\rho s},
\end{split}
\end{equation}
where $\{\chi^{A}\}$ is a minimal basis set (MBS) of atomic orbitals on the emitter atom and $D_{\mu r}$ are the expansion coefficients of the molecular orbitals in this basis.
In the case of the core orbital, $D_{\mu c} \simeq \delta_{\mu c}$.

The coefficients $D_{\mu r}$ are obtained by projecting the molecular orbitals onto the MBS.
Let $\{f_\kappa\}$ be the contracted Gaussian-type orbital (CGTO) basis in which the molecular orbital coefficients $C_{\kappa r}$ are defined ($\phi_r = \sum_\kappa f_\kappa C_{\kappa r}$).
Given the overlap matrix $T_{\mu\nu} = \langle \chi_\mu | \chi_\nu \rangle$ of the MBS and the overlap $U_{\mu\nu} = \langle \chi_\mu | f_\nu \rangle$ between the MBS and the CGTO basis, the projection yields
\begin{equation}\label{eqn:mbs_proj}
\mathbf{D} = \mathbf{T}^{-1} \mathbf{U} \mathbf{C}.
\end{equation}
Following Ref.~\cite{tenorio_multireference_2022}, in general, one selects the first fully contracted functions of the GTO basis as the MBS.
Since we use the STO-3G basis, the MBS is simply the subset of basis functions localized on the emitter atom.

The atomic two-electron integrals $\langle \chi_{Elm} \chi_c | \chi_\nu \chi_\rho \rangle$ in Eq.~\eqref{eqn:oca_approx} can be decomposed into radial integrals and analytical angular coefficients~\cite{mcguire_kshell_1969, walters_nonrelativistic_1971}.
We use the precomputed atomic integrals from the OpenMolcas package~\cite{galvan_openmolcas_2019, tenorio_multireference_2022}, which tabulate the fully assembled values of $\langle \chi_{Elm} \chi_c | \chi_\nu \chi_\rho \rangle$ for each pair of valence atomic orbitals and each partial wave $(l, m)$ of the continuum electron.

\section{Results and Discussion}

\subsection{System initialization}
To demonstrate the viability of our method, we calculate the Auger spectrum of a water molecule.
The Cartesian coordinates used in this work are listed in Appendix~\ref{app:geometries}.
We use the STO-3G basis set, yielding seven molecular orbitals (one core and six valence) and ten electrons.
In the GQE ground-state calculation, the oxygen 1s core orbital is frozen and excluded from the qubit register. The remaining six valence orbitals and eight electrons are mapped to 12 qubits.

At the q-sc-EOM stage, the core orbital is reintroduced into the active orbital space, allowing core-level ionizations (CVS-IP) and their associated Auger transitions to be described, expanding the system to 14 qubits.
The Hartree--Fock solution is used for the core orbital, while the GQE-optimized unitary operator $U$ acts on the valence space.
We exploit the $\mathrm{C}_{2\mathrm{v}}$ point-group symmetry of water to classify the excitation operators by irreducible representation (irrep), block-diagonalizing the $\mathbf{M}$ matrix as described in Sec.~\ref{subsec:qsceom}.
In the OCA calculations, the initial state is always the core-ionized state with a 1s hole on oxygen.
Accordingly, the spectra discussed below correspond to normal O $K-LL$ Auger decay in water, ending in doubly ionized valence states.

\subsection{GQE results}

\begin{figure}[htbp]
    \includegraphics[width=\columnwidth]{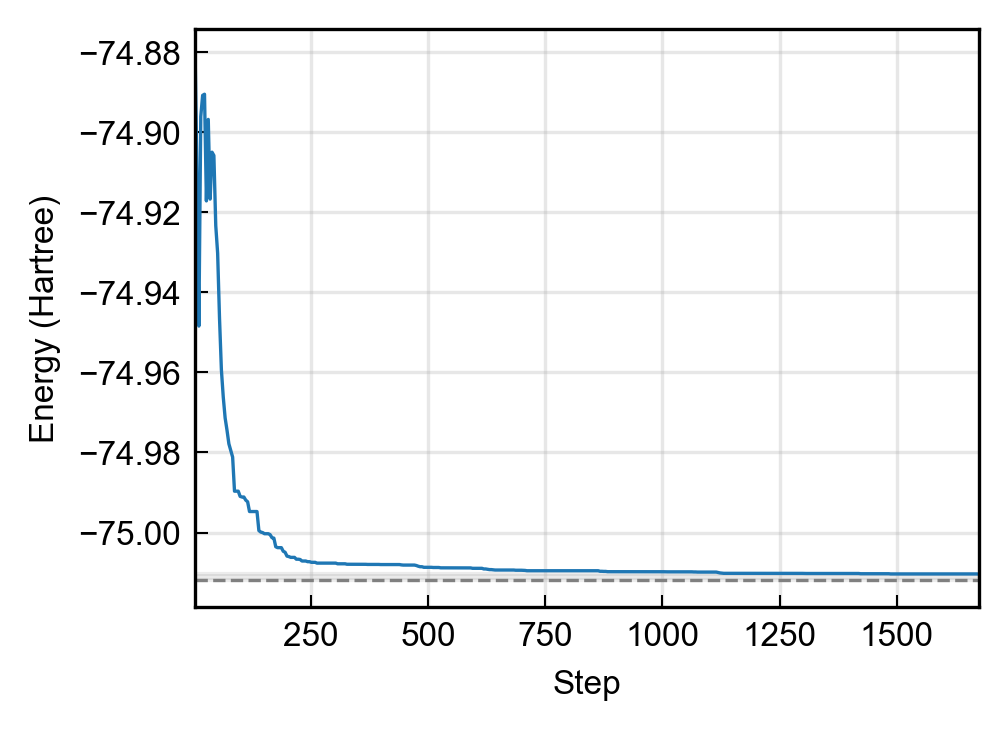}
    \caption{GQE results for the 12-qubit water molecule. The blue solid line is the GQE result, and the dashed grey line is the FCI result obtained using PySCF and the STO-3G basis. The grey region indicates the threshold for chemical accuracy. Simulation data are plotted only until this threshold is reached.}
    \label{fig:gqe-water}
\end{figure}

We set the GQE depth parameter $d$ to 60 tokens.
Chemical accuracy is achieved within 2,000 iterations of the GPT training cycle.
On a commercially available NVIDIA RTX A5000 GPU, results are obtained in less than 10 hours. The results of the ground-state calculation via GQE are presented in Fig.~\ref{fig:gqe-water}.

To assess circuit efficiency at fixed problem size, we compared frozen-core ground-state circuits for H$_2$O using first-order Trotterized  UCCSD-VQE and GQE under identical noiseless state-vector simulation settings in double precision.
The comparison is summarized in Table~\ref{tab:vqe-gqe-resource}.
At the same 12-qubit/8-electron setting, GQE reduces the total gate count by 58.8\% and the CNOT count by 64.2\% relative to UCCSD-VQE, while the energy is higher by 1.13 mHa.

\begin{table}[htbp]
  \caption{Ground-state resource comparison for frozen-core H$_2$O (STO-3G).}
  \label{tab:vqe-gqe-resource}
  \centering
  \begin{tabular}{lcc}
    \hline
    Metric & VQE (UCCSD) & GQE \\
    \hline
    Energy (Ha) & $-$75.0117639 & $-$75.0106362 \\
    $E - E_{\mathrm{VQE}}$ (mHa) & 0.000 & 1.128 \\
    Total gates & 2,628 & 1,082 \\
    CNOT gates & 1,504 & 538 \\
    \hline
  \end{tabular}
\end{table}

We also estimated the measurement workload required in the post-ground-state stages by counting the number of expectation-value evaluations implied by our implemented matrix-construction procedures.
For each symmetry block $\Gamma$, the $\mathbf{M}$ matrix workload is
\begin{equation}
N_{\mathrm{eval},M}^{(\Gamma)} = n_{\Gamma} + 2\binom{n_{\Gamma}}{2} = n_{\Gamma}^{2},
\end{equation}
where $n_{\Gamma}$ is the number of excitations in that block.
For Auger transition-RDM elements $R_{KI;csr}$, each operator component $\hat{a}_c^\dagger \hat{a}_s \hat{a}_r$ requires two Hermitian cross-block reconstructions between the IP and DIP bases, giving
\begin{equation}
N_{\mathrm{eval},R}^{(\Gamma)} = N_{csr}^{(\Gamma)} \times 2\left(n_{\mathrm{IP}}+n_{\mathrm{DIP}}^{(\Gamma)} + 2\,n_{\mathrm{IP}}\,n_{\mathrm{DIP}}^{(\Gamma)}\right).
\end{equation}
Here, the first two terms account for diagonal expectations needed by the superposition extraction, while the last term counts the cross-block off-diagonal measurements at two phase angles.
$N_{csr}^{(\Gamma)}$ denotes the number of Auger operator components $\hat{a}_c^\dagger \hat{a}_s \hat{a}_r$ that pass symmetry and spin selection for the target DIP irrep $\Gamma$.
For the same H$_2$O setting, the IP/DIP $\mathbf{M}$ matrix evaluations require 3,343 expectation-value evaluations in total, while the Auger transition-RDM evaluations require 43,896 expectation-value evaluations.
Thus, the combined total is 47,239 expectation-value evaluations, with the transition RDM step contributing about 92.9\% of the measurement workload.
The irrep-resolved breakdowns are summarized in Appendix~\ref{app:measurement-breakdown}.
We emphasize here that repeated use of the ground state unitary operator in the q-sc-EOM and OCA calculations means the shallower GQE reference (Table~\ref{tab:vqe-gqe-resource}) yields accumulated savings throughout the Auger spectrum calculation.

\subsection{Auger spectrum results}
\begin{figure}
    \begin{overpic}[width=\linewidth]{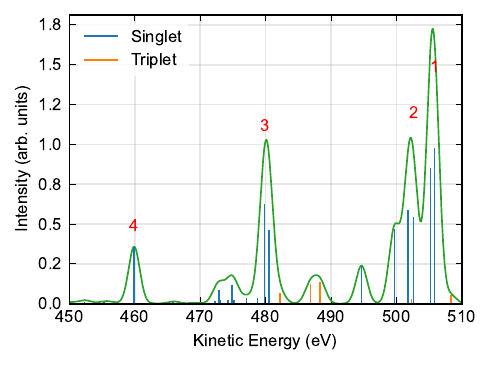}
        \put(1,72){\textbf{(a)}}
    \end{overpic}\\[0pt]
    \begin{overpic}[width=\linewidth]{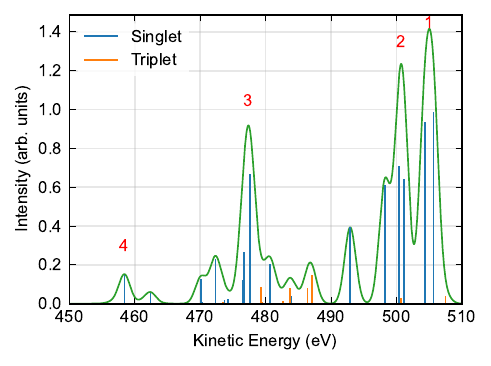}
        \put(1,72){\textbf{(b)}}
    \end{overpic}\\[0pt]
    \begin{overpic}[width=\linewidth]{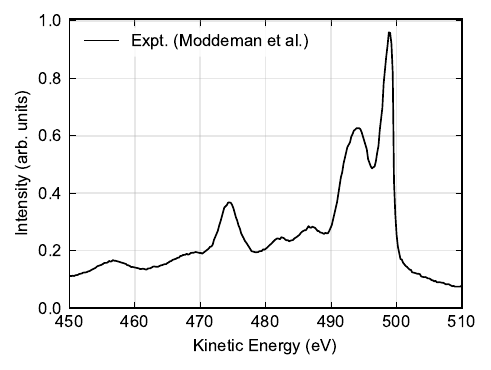}
        \put(1,72){\textbf{(c)}}
    \end{overpic}
    \caption{Auger spectrum of $\mathrm{H_2O}$. (a)~Spectrum obtained with the proposed hybrid workflow. (b)~Fully classical FCI+OCA reference. (c)~Experimental spectrum digitized from Ref.~\cite{moddeman_determination_1971}. We observe close agreement in relative peak intensity and position between all three spectra.}
    \label{fig:h2o_auger}
\end{figure}

Using the ground state obtained via GQE, we then apply the workflow described in Sec.~\ref{sec:methods} to calculate the Auger spectrum of water.
In all calculated spectra, the stick spectrum is convolved with a Gaussian envelope with a half-width at half-maximum (HWHM) of 1.0~eV.
The results of our calculations are displayed in Fig.~\ref{fig:h2o_auger}(a), along a classical reference generated from FCI in Fig.~\ref{fig:h2o_auger}(b) and an experimental spectrum digitized from Ref.~\cite{moddeman_determination_1971} in Fig.~\ref{fig:h2o_auger}(c).

We first compare the spectrum obtained via our workflow with one obtained classically by using FCI to calculate all excited states, followed by the OCA to compute the Auger spectrum.
Both spectra are similar in relative peak intensity and position when it comes to the most intense features, especially the two tallest peaks around 500--510 eV and the third-tallest peak around 480 eV.
However, the FCI result exhibits additional splittings, particularly around 480 eV and 450 eV, that are absent in the GQE spectrum. Since FCI is exact within the chosen basis set, these discrepancies likely originate from both the q-sc-EOM approximation and the ground-state error of GQE.

We also observe good agreement between our calculated and experimental spectra in relative peak positions and intensities.
In particular, the relative sizes and positions of the two largest peaks around 500 eV are very similar in both spectra.
These peaks appear at slightly lower energy in the experiment, just below 500 eV, whereas they appear between 500 and 510 eV in our calculation.
Both spectra also exhibit a prominent mid-energy peak with two smaller peaks on the higher-energy side.
In the experiment, this feature appears just above 470 eV, whereas in our calculation, it appears closer to 480 eV.
The two smaller peaks between the high-energy doublet and the single mid-energy peak are also present in both spectra.
Finally, in the low-energy regime below 460 eV, both spectra contain a single peak, although its relative intensity is higher in the calculated spectrum.

\begin{figure}
    \centering
    \includegraphics[width=\linewidth]{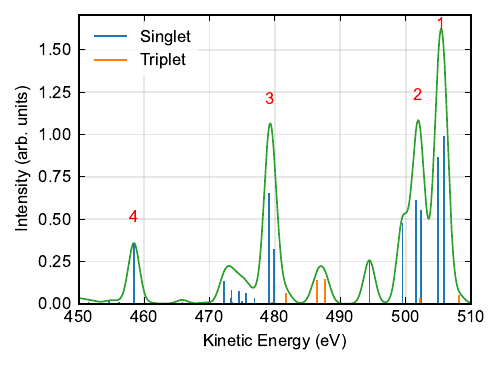}
    \caption{The Auger spectrum of water calculated using our workflow, but substituting VQE for GQE. This spectrum is nearly identical to the one produced in Fig.~\ref{fig:h2o_auger}(a), confirming that the GQE ground state is of comparable quality to VQE and that the differences from the FCI reference (Fig.~\ref{fig:h2o_auger}(b)) originate from the q-sc-EOM approximation for excited states.}
    \label{fig:vqe-auger}
\end{figure}

To further probe the accuracy of GQE, we compute the Auger spectrum using our hybrid workflow, but substitute VQE for GQE when calculating the ground state.
The result of this calculation is shown in Fig.~\ref{fig:vqe-auger}.
The spectrum obtained with VQE combined with the q-sc-EOM is nearly identical to that obtained with GQE and q-sc-EOM, confirming that the ground states of GQE and VQE are of comparable quality.
This also provides further supporting evidence that the differences between our result in Fig.~\ref{fig:gqe-water}(a) and the FCI result in Fig.~\ref{fig:gqe-water}(b) do not originate from GQE.

\subsection{Auger decay channel analysis}
To quantify the accuracy of our methods, we calculate the decay channels for doubly ionized water.
The corresponding numerical energies and relative intensities, together with the RASSCF/RASPT2 and experimental reference values, are summarized in Table~\ref{tab:auger-comparison}.
Based on these channels, we identify three characteristic regions of the Auger spectrum for water.
The first region spans approximately 493--510 eV and corresponds to channels involving a double outer-valence hole.
The strongest signals in this region come from the $1b_1^{-2}$ singlet channel at 505.83 eV (labelled 1 in Figs.~\ref{fig:h2o_auger} and~\ref{fig:vqe-auger}) and the $3a_1^{-2}$ singlet channel at 502.63 eV (labelled 2).
The second region spans approximately 460--489 eV.
In this region, the dominant signal comes from the $2a_1^{-1}3a_1^{-1}$ singlet channel at 479.84 eV (labelled 3).
The region also contains several other decay channels involving one outer-valence hole and one inner-valence hole.
We therefore denote this region as the outer-inner valence region.
Finally, the third region spans 450--460 eV and corresponds to the $2a_1^{-2}$ double-inner-valence-hole channel, with a signal at 459.88 eV (labelled 4).
The experimental spectrum, as well as the VQE spectrum, shows similar regions and the same major peaks in Figs.~\ref{fig:h2o_auger} and~\ref{fig:vqe-auger}.

Lastly, we compare these decay channels to those computed by VQE, those reported in the computational literature~\cite{tenorio_multireference_2022}, and those obtained via experiment~\cite{Inhester2012} in Table~\ref{tab:auger-comparison}.
We then match these channels and their Auger signatures with those obtained in our calculations using both GQE and VQE.
We find good agreement between our results and both references in relative peak locations, with the major labelled peaks 1, 2, 3, and 4 indicated by superscripts in Table~\ref{tab:auger-comparison}.
In both the GQE and VQE results, the signals associated with the $2a_1^{-2}$ and $2a_1^{-1}3a_1^{-1}$ configurations are about twice as intense as those in the references. However, given the consistency between GQE and VQE, this is likely due to the q-sc-EOM and not inaccuracies in the ground state.
We speculate that future algorithmic improvements may allow us to close the approximately 10 eV systematic gap between experiment and GQE.
However, we emphasize that overall our method accurately reproduces the Auger spectrum decay channels and the qualitative spectrum of doubly ionized water.

\begin{table*}[htbp]
\centering
\caption{Comparison of Auger energies and relative intensities for the proposed GQE + q-sc-EOM + OCA workflow (denoted GQE) and the corresponding workflow substituting VQE for GQE (denoted VQE), against values from a fully classical RASSCF/RASPT2 + OCA method~\cite{tenorio_multireference_2022} and an experimental reference from Ref.~\cite{Inhester2012}. Superscripts correspond to the labelled peaks in Fig.~\ref{fig:h2o_auger}. Relative intensities are rounded to the nearest whole number. Dashes indicate that no corresponding signal was obtained.}
\label{tab:auger-comparison}
\begin{tabular}{ccccc@{\hspace{0.5cm}}c@{\hspace{0.5cm}} c@{\hspace{0.5cm}} c@{\hspace{0.5cm}} c}
\hline
GQE & VQE & RASSCF & Expt & Configuration & $\Gamma_{\mathrm{rel}}$ GQE & $\Gamma_{\mathrm{rel}}$ VQE & $\Gamma_{\mathrm{rel}}$ RASSCF & $\Gamma_{\mathrm{rel}}$ Expt \\
(eV) & (eV) & (eV) &  &  &  &  &  &  \\
\hline
499.70 & 499.61 & 491.61 & 492.36 & $3a_1^{-1}1b_2^{-1}$ (S) & 48 & 48 & 73 & 68 \\
508.30 & 508.17 & 499.93 & 500.67 & $3a_1^{-1}1b_1^{-1}$ (T) & 5 & 5 & 1 & 3 \\
$^1$ 505.83 & 505.88 & 498.65 & 499.39 & $1b_1^{-2}$ (S) & 100 & 100 & 100 & 100 \\
505.23 & 505.03 & 497.33 & 497.98 & $3a_1^{-1}1b_1^{-1}$ (S) & 87 & 88 & 98 & 92 \\
503.72 & 503.57 & 495.64 & 496.60 & $1b_2^{-1}1b_1^{-1}$ (T) & 0 & 0 & 0 & 0 \\
$^2$ 502.63 & 502.46 & 493.86 & 494.64 & $3a_1^{-2}$ (S) & 56 & 56 & 70 & 70 \\
501.79 & 501.63 & 493.95 & 494.68 & $1b_2^{-1}1b_1^{-1}$ (S) & 60 & 62 & 86 & 80 \\
502.33 & 502.23 & 493.82 & 494.63 & $3a_1^{-1}1b_2^{-1}$ (T) & 3 & 3 & 1 & 2 \\
494.66 & 494.55 & 486.54 & 487.45 & $1b_2^{-2}$ (S) & 24 & 26 & 52 & 55 \\
488.34 & 487.74 & 481.78 & 482.30 & $2a_1^{-1}1b_1^{-1}$ (T) & 14 & 15 & 14 & 25 \\
486.86 & 486.53 & 480.73 & 480.58 & $2a_1^{-1}3a_1^{-1}$ (T) & 13 & 13 & 10 & 22 \\
482.24 & 481.89 & 477.14 & 476.82 & $2a_1^{-1}1b_2^{-1}$ (T) & 7 & 6 & 6 & 12 \\
480.50 & -- & 476.37 & 475.76 & $2a_1^{-1}1b_1^{-1}$ (S) & 48 & -- & 15 & 39 \\
$^3$ 479.84 & 479.21 & 473.54 & 473.27 & $2a_1^{-1}3a_1^{-1}$ (S) & 64 & 65 & 37 & 47 \\
-- & -- & 469.26 & 468.75 & $2a_1^{-1}1b_2^{-1}$ (S) & -- & -- & 11 & 26 \\
$^4$ 459.88 & 458.62 & 456.21 & 457.19 & $2a_1^{-2}$ (S) & 36 & 36 & 16 & 18 \\
\hline
\end{tabular}
\end{table*}

\section{Conclusion}
In this work, we present a computational workflow for calculating Auger spectra of molecules from molecular geometries using the GQE for accurate ground-state preparation, the q-sc-EOM method for excited-state and ionized-state calculations, and the OCA for Auger transition rates.
We demonstrate the validity of our method by calculating the Auger spectrum of water using the STO-3G basis set, mapped to 12 qubits for the ground-state calculation and 14 qubits for the excited-state and Auger calculations.
We successfully reproduce the number as well as relative positions and intensities of the major peaks in three distinct spectral regions: the double outer-valence hole region (493--510 eV), the outer-inner valence region (460--489 eV), and the double inner valence region (450--460 eV).
A quantitative comparison of individual decay channels with RASSCF/RASPT2 + OCA calculations and experimental data confirms good agreement in the Auger kinetic energies.
We also find that GQE halves the total gate count compared to VQE, providing an amortized resource advantage.

Potential future work includes using more sophisticated basis sets, further hyperparameter tuning of GPT-2, and extending GQE to generate excited-state circuits directly rather than relying on the q-sc-EOM.
Currently, we use the STO-3G basis. However, recent work by Skomorowski et al. has demonstrated calculation of accurate Auger spectra for molecules as large as benzene using the cc-pVTZ basis set~\cite{Skomorowski2025OCA}.
Using a more complete basis set may yield similar gains in our workflow.
An alternative approach would be to abandon global-basis sets like STO-3G and cc-pVTZ altogether.
Knottmann et al. have demonstrated global-basis free quantum molecular descriptions can provide highly accurate results while significantly reducing the qubit cost~\cite{reducing-qubit}.
Furthermore, investigating the effect of the classical optimizer and its hyperparameters on GQE convergence could further reduce the computational cost of ground-state preparation.
Finally, extending GQE to directly generate excited-state circuits (rather than relying on q-sc-EOM) and establishing molecule-to-molecule transferability would allow the full workflow to benefit from the advantages of GPU-HPC systems.
Additional error mitigation techniques will be necessary to implement this algorithm on real devices.

These improvements, combined with the scalability of the GQE framework driven by GPU-accelerated transformer models and the potential for transfer learning across molecular systems, make GQE well-positioned to simulate molecules larger than 40 qubits, the current limit of classical FCI~\cite{sun_recent_2020, gao_distributed_2024}.
Small organic and organometallic compounds used in EUV photoresist design are an industrially relevant first target.
The ability to accurately calculate Auger spectra for large systems will accelerate the development of next-generation materials for semiconductor manufacturing and beyond.

\section{Acknowledgements}
A part of this work was performed for Council for Science, Technology and Innovation (CSTI), Cross-ministerial Strategic Innovation Promotion Program (SIP), “Promoting the application of advanced quantum technology platforms to social issues” (Funding agency: QST).
Some of the computations were performed using the high-performance computing (HPC) system NAYUTA of Mitsubishi Chemical Corporation (MCC). “NAYUTA” is an alias for the MCC HPC system and is not the name of a product or service provided by MCC. 
Some of the results presented in this paper were obtained using AIST G-QuAT's ABCI-Q.

Y.N. gratefully acknowledges support from the NSERC CREATE for Accelerated Discovery (AccelD) training program hosted by the Acceleration Consortium (Grant \#596133-2025) and was partially supported through a collaborative partnership with Moderna Inc. A.A.-G. thanks Anders G. Fr{\o}seth for his generous support. A.A.-G. also acknowledges the generous support of Natural Resources Canada and the Canada 150 Research Chairs program. This research is part of the University of Toronto’s Acceleration Consortium, which receives funding from the CFREF-2022-00042 Canada First Research Excellence Fund.

\newpage
\appendix

\section{XAS for Lithium Hydride} \label{app:lih-xas}

As a complementary test of our workflow, we calculate the X-ray absorption spectroscopy (XAS) spectrum of LiH.
XAS probes core-level excitations and provides a test case for the transition density matrix calculation.
We use the STO-3G basis set, yielding six molecular orbitals (one core and five valence) and four electrons.

In the GQE ground-state calculation, the lithium 1s core orbital is frozen and excluded from the qubit register. The remaining five valence orbitals and two electrons are mapped to 10 qubits.
We find that a circuit depth of 20 tokens is sufficient to reach chemical accuracy in the ground state using GQE. However, there may be room for further optimization through parameter tuning of the optimizer, which is beyond the scope of this study.
At the q-sc-EOM stage, the core orbital is reintroduced with its Hartree--Fock solution, and the point-group symmetry is reduced to $\mathrm{C}_{2\mathrm{v}}$ to classify the excitation operators by irreducible representation and block-diagonalize the $\mathbf{M}$ matrix.

We target core excitations using particle-number-conserving excitation operators $\hat{G}_u \in \{ \hat{a}^\dagger_a \hat{a}_c, \, \hat{a}^\dagger_a \hat{a}^\dagger_b \hat{a}_i \hat{a}_c \}$ with $c$ denoting a core orbital, where a CVS constraint retains only excitations with exactly one core orbital annihilation index; double core-hole excitations are not considered.
The core-excited energies $E_n^{\mathrm{XAS}}$ are obtained by diagonalizing the $\mathbf{M}$ matrix (Eq.~\eqref{eqn:eig}) constructed from these operators.

The XAS spectrum is computed from the one-body transition RDM between the ground state and each core-excited eigenstate,
\begin{equation}\label{eqn:trdm}
\gamma_{pq}^{(n)} = \langle \Psi_0 | \hat{a}_p^\dagger \hat{a}_q | \Psi_n \rangle,
\end{equation}
obtained by setting $\hat{T} = \hat{a}_p^\dagger \hat{a}_q$ for each pair $(p,q)$.
The transition dipole moment for the $0 \to n$ transition is
\begin{equation}\label{eqn:trdipole}
\mu_{0n}^{(\alpha)} = \sum_{p,q} \gamma_{pq}^{(n)} \, d_{pq}^{(\alpha)}, \qquad \alpha = x, y, z,
\end{equation}
where $d_{pq}^{(\alpha)} = \langle p | r_\alpha | q \rangle$ are the one-electron dipole integrals in the molecular orbital basis.
The isotropic oscillator strength is then given by
\begin{equation}\label{eqn:osc_strength}
f_{0n} = \frac{2}{3} \Delta E_{0n} \sum_{\alpha = x, y, z} \left| \mu_{0n}^{(\alpha)} \right|^2,
\end{equation}
where $\Delta E_{0n} = E_n - E_0$ is the excitation energy. 

\begin{figure}[H]
    \includegraphics[width=\columnwidth]{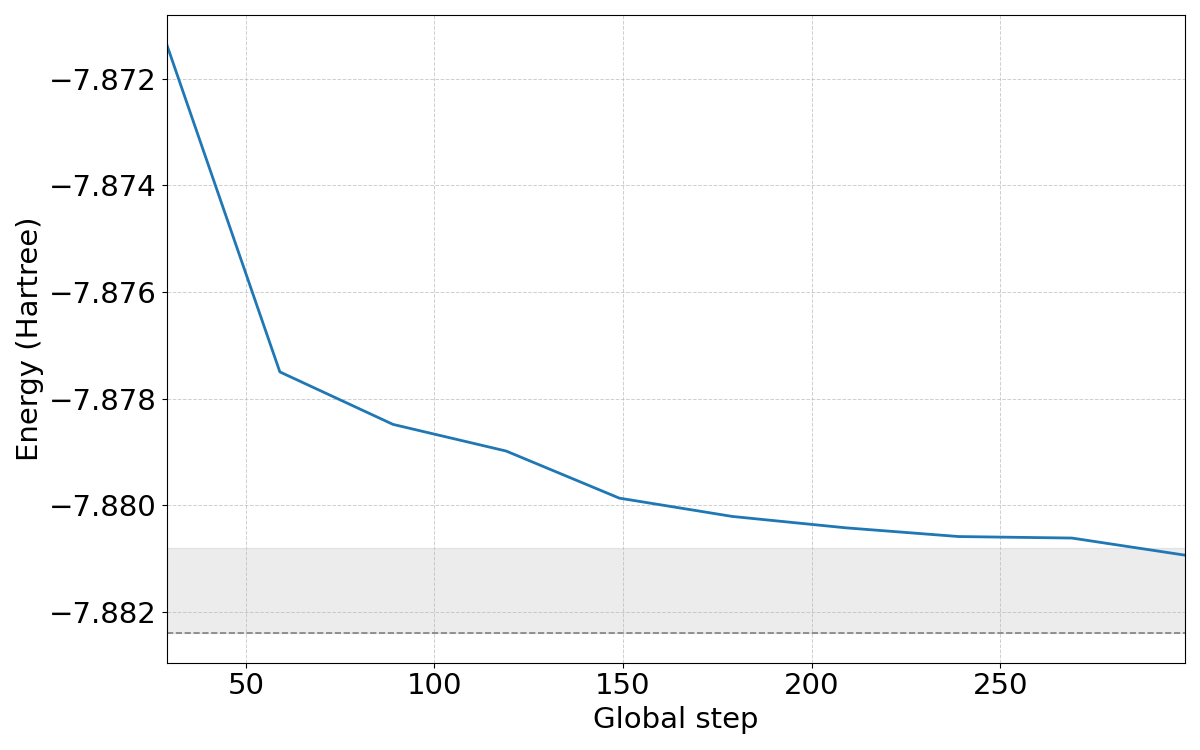}
    \caption{GQE results for the LiH model. The solid blue line is the GQE result, and the dashed grey line is the FCI result obtained by PySCF. The solid grey area indicates the range of chemical accuracy. The result obtained by GQE is within chemical accuracy of the FCI result.}
    \label{fig:gqe-lih}
\end{figure}

\begin{figure}[H]
    \includegraphics[width=\columnwidth]{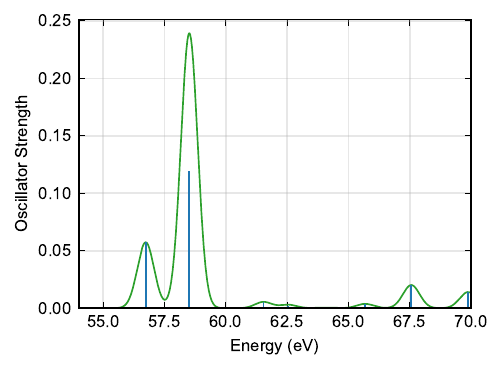}
    \caption{X-ray absorption spectrum for LiH. The stick spectrum is convolved with a Gaussian envelope with HWHM = 0.4 eV.}
    \label{fig:lih-xas}
\end{figure}

\section{Cartesian coordinates} \label{app:geometries}

Table~\ref{tab:h2o-coordinates} and Table~\ref{tab:lih-coordinates} list the Cartesian coordinates used in the H$_2$O Auger and LiH XAS calculations, respectively.
All coordinates are given in angstrom.

\begin{table}[htbp]
  \caption{Cartesian coordinates used for H$_2$O.}
  \label{tab:h2o-coordinates}
  \centering
  \begin{tabular}{lrrr}
    \hline
    Atom & $x$ & $y$ & $z$ \\
    \hline
    H & $-0.7586$ & $0.0000$ & $0.5870$ \\
    O & $0.0000$ & $0.0000$ & $0.0000$ \\
    H & $0.7586$ & $0.0000$ & $0.5870$ \\
    \hline
  \end{tabular}
\end{table}

\begin{table}[htbp]
  \caption{Cartesian coordinates used for LiH.}
  \label{tab:lih-coordinates}
  \centering
  \begin{tabular}{lrrr}
    \hline
    Atom & $x$ & $y$ & $z$ \\
    \hline
    Li & $0.0000$ & $0.0000$ & $0.0000$ \\
    H & $0.0000$ & $0.0000$ & $1.6000$ \\
    \hline
  \end{tabular}
\end{table}

\section{Irrep-resolved measurement workload breakdown} \label{app:measurement-breakdown}
Table~\ref{tab:appendix-m-obs} and Table~\ref{tab:appendix-trs-obs} provide the irrep-resolved measurement workload breakdowns.
These irrep-resolved counts were obtained from the same procedure for counting measurements used for the main text estimates.

\begin{table}[htbp]
  \caption{Irrep-resolved excitation counts and estimated evaluation count for IP/DIP M matrices.}
  \label{tab:appendix-m-obs}
  \centering
  \begin{tabular}{lrrrr}
    \hline
    Irrep & $n_{\mathrm{IP}}$ & $N_{\mathrm{eval,IP}}$ & $n_{\mathrm{DIP}}$ & $N_{\mathrm{eval,DIP}}$ \\
    \hline
    $\mathrm{A}_1$ & 10 & 100 & 30 & 900 \\
    $\mathrm{A}_2$ & 3 & 9 & 26 & 676 \\
    $\mathrm{B}_1$ & 3 & 9 & 28 & 784 \\
    $\mathrm{B}_2$ & 9 & 81 & 28 & 784 \\
    \hline
    Total & 25 & 199 & 112 & 3144 \\
    \hline
  \end{tabular}
\end{table}

\begin{table}[htbp]
  \caption{Irrep-resolved transition RDM workloads for Auger $R_{KI;csr}$. The selected IP sector is $\mathrm{A}_1$ with $n_{\mathrm{IP}}=10$.}
  \label{tab:appendix-trs-obs}
  \centering
  \begin{tabular}{lrrrr}
    \hline
    DIP irrep & $n_{\mathrm{DIP}}$ & $n_{\mathrm{IP}}+n_{\mathrm{DIP}}+2\,n_{\mathrm{IP}}\,n_{\mathrm{DIP}}$ & $N_{csr}$ & $N_{\mathrm{eval},R}$ \\
    \hline
    $\mathrm{A}_1$ & 30 & 640 & 14 & 17920 \\
    $\mathrm{A}_2$ & 26 & 556 & 4 & 4448 \\
    $\mathrm{B}_1$ & 28 & 598 & 6 & 7176 \\
    $\mathrm{B}_2$ & 28 & 598 & 12 & 14352 \\
    \hline
    Total & 112 & -- & 36 & 43896 \\
    \hline
  \end{tabular}
\end{table}

\bibliography{bibliography}

\end{document}